\documentclass[aps,prd,preprint,superscriptaddress,showpacs,nofootinbib]{revtex4}

\usepackage{latexsym}
\usepackage{amssymb}
\usepackage{epsfig,amsmath,graphics}
\usepackage{pstricks}
\usepackage{color}
\usepackage{dcolumn}

\newcommand{\gev}{\text{GeV}}

\newcommand{\vev}[1]{{\langle #1 \rangle}}
\newcommand{\dcg}{\delta c_{\gamma \gamma}}
\newcommand{\dcv}{\delta c_V}
\newcommand{\dcf}{\delta c_f}

\usepackage{comment}
\newcommand{\be}{\begin{equation}}
\newcommand{\ee}{\end{equation}}

\newcommand{\Mev}{{\rm MeV}}

\def\bea{\begin{eqnarray}}
\def\eea{\end{eqnarray}}
\def\ltap{\ \raise.3ex\hbox{$<$\kern-.75em\lower1ex\hbox{$\sim$}}\ }
\def\gtap{\ \raise.3ex\hbox{$>$\kern-.75em\lower1ex\hbox{$\sim$}}\ }
\def\lsim{\ \raise.3ex\hbox{$<$\kern-.75em\lower1ex\hbox{$\sim$}}\ }
\def\gsim{\ \raise.3ex\hbox{$>$\kern-.75em\lower1ex\hbox{$\sim$}}\ }

\usepackage{slashed}

\usepackage{color}

\begin{document}

\preprint{FERMILAB-PUB-12-401-T}

\title{Higgs Signals in a Type I 2HDM or with a Sister Higgs}

\author{Daniele S. M. Alves$^{a}$, Patrick J. Fox$^{a}$ and Neal J. Weiner$^{b}$\\
$^a$ Theoretical Physics Department, \\Fermi National Accelerator Laboratory, Batavia, IL 60510, USA\\
$^b$ Center for Cosmology and Particle Physics, \\Department of Physics, New York University, New York, NY 10003}

\begin{abstract}
In models where an additional SU(2)-doublet that does not have couplings to fermions participates in electroweak symmetry breaking, the properties of the Higgs boson are changed. At tree level, in the neighborhood of the SM-like range of parameter space, it is natural to have the coupling to vectors, $c_V$, approximately constant, while the coupling to fermions, $c_f$, is suppressed. This leads to enhanced VBF signals of $\gamma \gamma$ while keeping other signals of Higgs production approximately constant (such as $WW^*$ and $ZZ^*$), and suppressing $h\rightarrow \tau \tau$. Sizable tree-level effects are often accompanied by light charged Higgs states, which lead to important constraints from $b\rightarrow s \gamma$ and $t \rightarrow b H^+$, but also often to similarly sizable contributions to the inclusive $h\rightarrow \gamma \gamma$ signal from radiative effects. In the simplest model, this is described by a Type I 2HDM, and in supersymmetry is naturally realized with ``sister Higgs'' fields. In such a scenario, additional light charged states can contribute further with fewer constraints from heavy flavor decays. With supersymmetry, Grand Unification motivates the inclusion of colored partner fields. These G-quarks may provide additional evidence for such a model.
\end{abstract}

 \maketitle

\section{Introduction}
Recently, both the ATLAS \cite{ATLAS:2012ad} and CMS \cite{Chatrchyan:2012tw} experiments have reported signals in $\gamma \gamma$ that are consistent with a Higgs signal with $m_h = 125\, \gev$ - enough to claim a discovery of a new particle, which is likely the Higgs. While the signs are quite tentative, there are also indications in the data that - if real - the couplings of the Higgs may not be consistent with the standard model. In particular, while the overall $h \rightarrow \gamma \gamma$ signal is slightly high, the vector boson fusion $h\rightarrow \gamma \gamma$ is possibly quite enhanced at CMS \cite{Chatrchyan:2012tw,Giardino:2012ww,Giardino:2012dp}. At the same time, the diboson decays ($WW^*/ZZ^*$) seem in line with the standard model. The only sign of fermionic decays - to $\tau \tau$ - has so far not been conclusively observed.

Although the data remain ambiguous, with the significant new data expected in 2012 alone, it is worth taking the possibility of a non-SM Higgs seriously. While some have pursued phenomenological studies \cite{Giardino:2012dp}, one should consider what sorts of models might actually naturally lead to these deviations, and, indeed, numerous proposals have appeared \cite{Ferreira:2012my,Buckley:2012em,Cao:2012yn,Cohen:2012wg,Abe:2012fb,Joglekar:2012hb,Bertolini:2012gu,ArkaniHamed:2012kq}. Enhancements of $h\rightarrow \gamma \gamma$ are difficult, as matter fields generally destructively interfere with the existing signal (although see \cite{Bai:2011aa,Dobrescu:2011aa}). Much effort has gone into studying the possible signals in the MSSM \cite{Djouadi:2005gj,Carena:2011aa,Carena:2012xa,Carena:2012gp} and in Type II 2HDM and 2HDM generally \cite{Ferreira:2011aa,Blum:2012kn}. In such models, the dominant effect arises from suppressions of the width to $b \bar b$, which allows the decays to $\gamma \gamma$ to dominate. At the same time, this would enhance {\em all} $\gamma \gamma$ signals, not preferably the VBF. At present there is also no dramatic signal of $W W^*$, which would also be enhanced by the same. 

An alternative possibility that we shall pursue here is to consider an additional Higgs $H$ that does not couple to fermions, but does participate in electroweak symmetry breaking. The advantage of such a new field is that it is manifestly quite safe, as it does not introduce new sources of flavor violation, or large interactions with SM fermions. In the most limited case of the SM+H, this scenario reduces to a Type I 2HDM. More generally, such a particle has been dubbed a ``sister Higgs'' \cite{sh}, particularly in the context of supersymmetric theories. While the branching ratios of the Higgs in such models has been studied before (e.g., \cite{Branco:2011iw,Posch:2010hx}), it is clear what is really interesting is the overall scaling of the different signals ($\sigma \times {\rm BR}$), which shall be our focus. We shall include both tree and loop level corrections\cite{Spira:1995rr}, as well as impose the non-trivial recent constraints on $t\rightarrow b H^+$. As we shall see,  $gg\rightarrow h\rightarrow \gamma \gamma$ is often enhanced, while VBF, associated production (WH/ZH/$t \bar t h$/$b \bar b h$) with $h\rightarrow \gamma \gamma$ are all enhanced even further. $g g \rightarrow h \rightarrow W W^*/Z Z^*$ is typically comparable to the SM while $g g \rightarrow h \rightarrow \bar \tau \tau$ is suppressed.

\subsection{The Couplings of the Higgs Boson}

The properties of the Higgs are essentially determined by its couplings to SM states (see discussion of general parameter space in \cite{Giardino:2012ww,Azatov:2012rd,Giardino:2012dp,Carmi:2012yp,Carmi:2012zd}).  The couplings: $c_V$ and $c_t$ determine its dominant production mechanisms: VBF ($c_V$), Vh ($c_V$), and gg fusion ($c_t$).  Its principle decays depend on these and other parameters: decays into vector bosons $c_V$, decays into photons $c_V,c_t$ and its decays into bottom quarks $c_b$. Additional small corrections to the width come from $c_c$ and an additional observable, namely the decay to $\bar \tau \tau$ comes from $c_\tau$.  

The most significant new physics processes that contribute to the properties of the Higgs come in the form of a new contribution to the $\gamma \gamma$ decay amplitude, $\delta c_{\gamma \gamma} \times \mathcal{M}^{SM}_{\gamma\gamma}$, and an additional contribution to the $gg$-fusion production amplitude, $\delta c_{gg}  \times \mathcal{M}^{SM}_{gg}$. We neglect the possibility of an additional invisible  contribution here.

With these parameters we have (for a 125 GeV Higgs),
\bea
\sigma_{gg} = (c_t+\delta c_{gg})^2 \sigma_{gg}^{SM},\\
\sigma_{VBF,VH} = c_V^2 \sigma_{VBF,VH}^{SM},\\
\sigma_{\bar f_i f_i h} = c_{f_i}^2 \sigma_{\bar f_i f_i h}^{SM},\\
\Gamma_{\gamma\gamma} = (1.28 c_V-0.28 c_t + \delta c_{\gamma\gamma})^2 \Gamma_{\gamma\gamma}^{SM},\\
\Gamma_{VV} = c_V^2 \Gamma_{VV}^{SM},\\
\Gamma_{\bar f_i f_i} = c_{f_i}^2 \Gamma_{\bar f_i f_i}^{SM}~.
\eea

Within this parameterization we have great freedom to consider models with different couplings to different generations, up and down-type fermions, leptons versus quarks, different suppressions to fermions versus gauge bosons, as well as contributions from charged and colored loops.  However, there is perhaps too much freedom to study this properly, and so it makes sense to focus on physically motivated scenarios. For instance, it makes sense to set $c_t=c_c=c_u$, and $c_b=c_s=c_d=c_\tau=c_\mu=c_e$ (i.e., a type II 2HDM, such as in supersymmetry.) Such a parameter space can lead to a enhancement of almost all signals through the suppression of the $\bar b b$ width, but will not, in general, preferentially boost the $\gamma \gamma$ signals compared to the e.g., $WW^*/ZZ^*$. 

Thus, instead, let us consider a Type I 2HDM.

\section{Signals from a Sister Higgs}
We use the notation of \cite{Mahmoudi:2009zx} and consider a general potential for the Higgs fields $\Phi_1$ and $\Phi_2$ both doublets with hypercharge $Y=1/2$, 
\bea
\nonumber{ \cal L} &=& D_\mu \Phi_1 (D^\mu  \Phi_1)^\dagger+D_\mu  \Phi_2 (D^\mu  \Phi_2)^\dagger-\left[ m_1^2  \Phi_1  \Phi_1^\dagger + m_2^2  \Phi_2  \Phi_2^\dagger + \frac{\lambda_1}{2}| \Phi_1|^4\right.\\ && + \frac{\lambda_2}{2}| \Phi_2|^4+\lambda_3 | \Phi_1|^2 | \Phi_2|^2 + \lambda_4 | \Phi_1^\dagger  \Phi_2|^2\\ && \nonumber  \left.+\left\{ \frac{\lambda_5}{2} ( \Phi_1^\dagger  \Phi_2)^2 + ( \Phi_1^\dagger  \Phi_2)(-m_{12}^2+\lambda_6 | \Phi_1|^2 + \lambda_7 | \Phi_2|^2) + h.c. \right\}\right]~.
\label{eq:higgspot}
\eea
For simplicity we assume throughout that CP is conserved and all parameters in (\ref{eq:higgspot}) are real.

As only one Higgs couples to fermions, we have the simplifying assumption that $c_t=c_b...=c_f$. As we are assuming that the theory is at least approximately SM-like, so it makes sense to perturb away from the SM values $c_v= 1 + \delta c_V$, $c_f = 1+\delta c_f$ and $\delta c_{\gamma \gamma}$ small. We define the ratio $R_i(X)$ as the ratio $\sigma_i(X)/\sigma_i^{SM}(X)$, for a production process $i$ and Higgs decay mode $X$. In this expansion we find 
\bea
R_{gg}(\gamma \gamma) = R_{t\bar{t}}(\gamma\gamma) &=& 1+ 2\, \dcg- 0.07\, \dcf + 2.1\, \dcv,\\
R_{VBF}(\gamma \gamma) = R_{Vh}(\gamma\gamma)  &=& 1+2\, \dcg - 2.07\, \dcf + 4.07\, \dcv,\\
R_{gg}(VV) = R_{VBF}(ff) = R_{Vh}(ff)  &=& 1-0.005\, \dcg + 0.5\, \dcf + 1.51\, \dcv,\\
R_{gg}(ff)  &=& 1-0.005\, \dcg + 2.5\, \dcf - 0.5\, \dcv,\\
R_{VBF}(VV)  &=& 1-0.005\, \dcg - 1.51\, \dcf + 3.5\, \dcv.
\eea

Two observations are immediately in order. First, and unsurprisingly, only $\gamma \gamma$ decays are very sensitive to $\dcg$. The second, and perhaps most important observation, is that $R_{gg}(\gamma \gamma) $ is only very weakly sensitive to $\dcf$ while all other modes are quite sensitive. This is simply because as $c_f$ is dialed down, the $gg$ fusion process is suppressed, but so too is the $b \bar b$ decay mode (boosting the $\gamma \gamma$ BR, such that the $\dcf$ dependence approximately cancels). The remaining increase in the $\gamma \gamma$ branching ratio from a reduction in the destructive interference of the top quark contribution leaves the overall sensitivity of $R_{gg}(\gamma \gamma)$ to $\dcf$ an order of magnitude below the other channels.  The upshot of this is quite striking: {\em in the neighborhood of the SM, for suppressed $c_f$, the $R_{gg}(\gamma \gamma)$ is quite stable, while decays to fermions will drop rapidly and $R_{VBF}(\gamma \gamma)$ will increase.}

It is worth going a step further to put this more into the language of the Higgs states.
The interactions of the low energy eigenstates can be written in terms of the angles $\tan \beta = \langle{\Phi_2}\rangle/\langle{\Phi_1}\rangle = v_2/v_1$, and $\alpha$, which diagonalizes the mass matrix of the CP-even Higgses.

The tree-level couplings of the light mass eigenstate to the gauge bosons and fermions are \cite{Branco:2011iw}
\bea
 c_f &=& \frac{v g_{h f \bar f}}{m_f}= \cos\alpha/\sin\beta\\
 c_V &=& \frac{v g_{h VV}}{2 m_V^2}=\sin(\beta-\alpha),
 \eea

It is well known that for $\alpha = \pi/2$, the theory reduces to a fermiophobic Higgs boson \cite{Branco:2011iw}. However, if we want this theory to describe the signal near 125 GeV reported by the CMS and ATLAS collaborations, it must be at least approximately SM-like. As $c_f=c_V=1$ when $\alpha=\beta-\pi/2$, it is more convenient to define $\delta = \beta - \alpha-\pi/2$ (i.e., so when $\delta=0$ the light mass eigenstate has SM-like couplings). Around this point 
\bea
c_f &=& \cos \delta- \cot \beta \sin \delta, \\
c_V &=& \cos\delta .
\eea
From this parameterization, we can already make two interesting observations: first while $c_V$ is constant at leading order in $\delta$, $c_f$ depends linearly on $\delta$. Second, this dependence of $c_f$ on $\delta$ is only sizeable for $\tan\beta \sim 1$. This simple point is one of the principle observations of this letter: that in the neighborhood of the SM Higgs, a Type I 2HDM at small $\tan \beta$ provides a simple realization of a scenario where $c_V \approx 1$ is nearly constant as a local function of $\delta$, but $c_f < 1$ and in general can deviate from one significantly. 

Combining this with our above discussion of the dependence of physical observables on $\dcf$ and $\dcv$, we can now discuss the tree-level (mixing) consequences of a Type I 2HDM on physical observables. The total width of the Higgs boson can be written as
\be
\Gamma_{tot} = c_f^2 \Gamma^{SM}_{f} + c_V^2 \Gamma^{SM}_{V}+\Gamma_{\gamma\gamma},
\ee
where $\Gamma^{SM}_f = \Gamma_{b\bar{b}}+ \Gamma_{c\bar{c}}+\ldots + \Gamma_{gg}$ (as $\Gamma_{gg}$ arises from a top loop, it also scales with $c_f^2$, we assume that $\delta c_{gg}=0$), and $\Gamma^{SM}_V=\Gamma_{ZZ}+\Gamma_{WW}$.  For the standard model (SM) higgs \cite{LHCHiggsxsecWGroup} at 125 GeV $\Gamma^{SM}_f =3.2\ \Mev$, $\Gamma^{SM}_V= 0.98\ \Mev$ and $\Gamma^{SM}_{\gamma\gamma}= 9.3\times 10^{-3}\ \Mev$.

\begin{figure}[t] 
   \centering
   \includegraphics[width=0.45\textwidth]{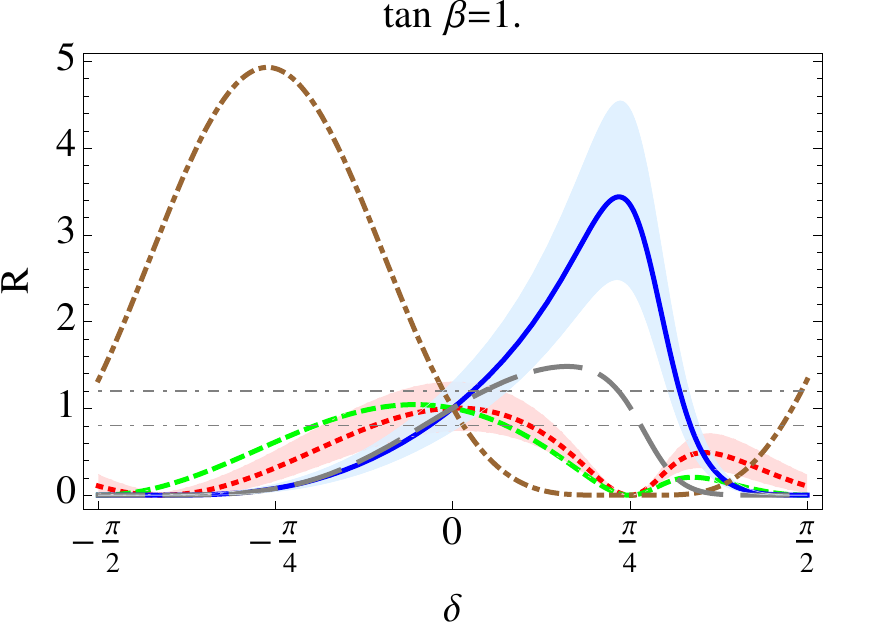} 
   \includegraphics[width=0.45\textwidth]{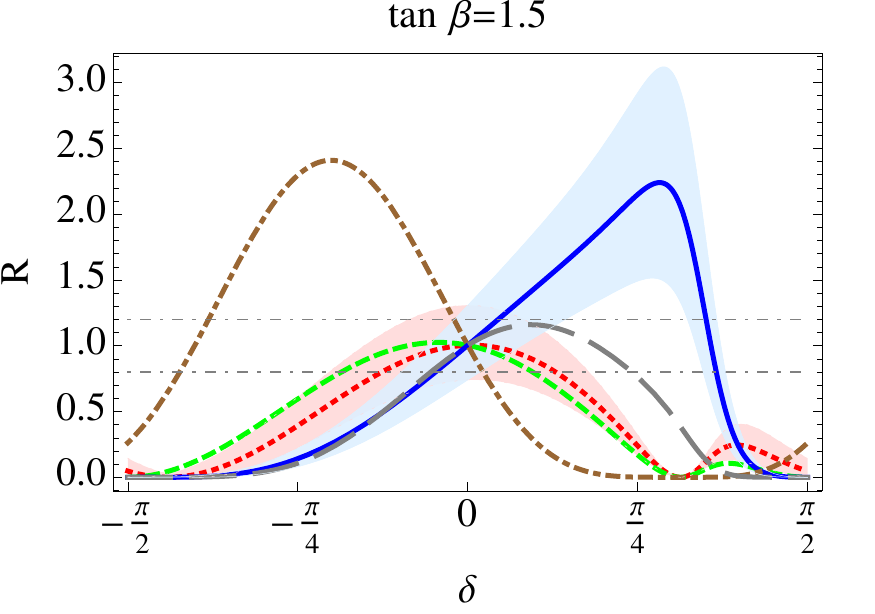} \\
   \includegraphics[width=0.45\textwidth]{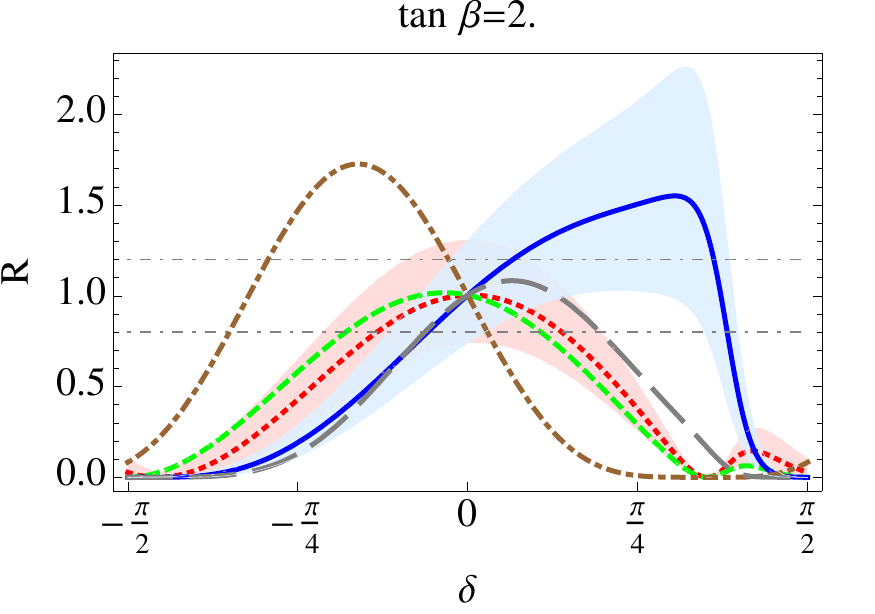} 
   \includegraphics[width=0.45\textwidth]{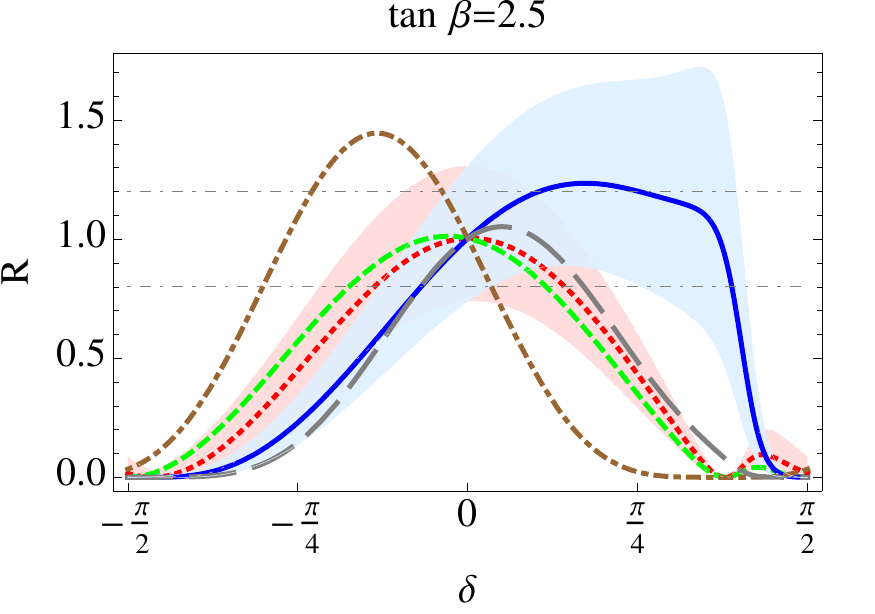} 
   \caption{$\sigma\times BR$, relative to the SM ($\delta=0$), for gluon fusion to $\gamma\gamma$ (dotted red), gluon fusion to $WW^*/ZZ^*$ (dashed green), VBF to $\gamma\gamma$ (solid blue), gluon fusion to $\tau\tau$ (dot-dashed brown), and VBF to $WW^*/ZZ^*$ (long-dashed gray).  The horizontal lines are at the SM rate $\pm 20\%$.  The lines ignore the contribution of new charged states to the $\gamma\gamma$ width.  The bands show the effect of an additional charged state whose loop contribution is smaller in magnitude than 50\% of the top contribution.}
   \label{fig:sigBRrel}
\end{figure}

Let us begin by ignoring the loop effects of additional charged states, i.e., we temporarily set $\delta c_{\gamma\gamma}=0$, and $\delta c_{gg}=0$. In this limit, we can focus on the tree-level effects exclusively. With only two remaining parameters determining the rates of these processes this Type I 2HDM setup is very simple to understand.  

\bea
R_{gg}(\gamma \gamma) = R_{t\bar{t}}(\gamma\gamma) &=& 1+0.07\, \delta \cot \beta - \delta^2 (1+0.7 \cot^2\beta),\\
R_{VBF}(\gamma \gamma) = R_{Vh}(\gamma\gamma)  &=&  1+2.1\, \delta \cot\beta+ \delta^2(2.5 \cot^2\beta-1),\\
R_{gg}(VV) = R_{VBF}(ff) = R_{Vh}(ff)  &=& 1 - 0.49\,\delta \cot\beta - \delta^2 (1+0.5\cot^2\beta),\\
R_{gg}(ff)  &=& 1-2.5\, \delta \cot\beta +\delta^2 (1.5\cot^2\beta-1),\\
R_{VBF}(VV)  &=& 1+1.5\,\delta\cot\beta +\delta^2 (1.53\cot^2\beta-1)~.
\eea
Note that for the standard gluon-fusion $\gamma \gamma$ signal, the linear term in $\delta$ is quite small, while for VBF $\gamma \gamma$ and gluon fusion to $WW^*/ZZ^*$ it is sizable, and with opposite signs. Thus, for positive delta, we expect in the vicinity of the SM-like Type I model, the inclusive $\gamma \gamma$ signal should not change significantly, while the VBF and associated production $\gamma \gamma$ signals should be enhanced, and the diboson $WW^*/ZZ^*$ decays are more moderatly suppressed. At larger positive (negative) $\delta$ the standard $\gamma \gamma$ signal will be suppressed (enhanced). We show the relative signals for the exact tree-level expressions as the solid central lines in Figure \ref{fig:sigBRrel}. 

Here we see how important the tree-level contributions can be to VBF and associated production signals. At small $\tan \beta$, in particular, we seem important deviations. For $\tan\beta = 1.5$, for instance, a 50\% boost in the VBF/associated production signals can occur all while keeping the diboson channels within 80\% of the SM values, while the ditau signal is simultaneously suppressed to near nothing. Of course, there are not only tree-level effects, but also loop level effects as well, and ignoring these could lead one to erroneous conclusions about the signals of such models. In the presence of a moderate radiative effect, corrections to the inclusive $\gamma \gamma$ signature can be significant as well.

In Type I 2HDMs there is a charged Higgs that can contribute to $\Gamma_{\gamma\gamma}$, making such corrections natural. However, unlike Type II models, where constraints such as $b \rightarrow s \gamma$ and $t \rightarrow b H^+$ are extremely strong, in this case they can be quite weak, with the charged Higgs becoming quite light $m_{H^+} \approx 100 \, \gev$. We shall discuss this shortly.

\section{A Parametric Study}

Unlike Type II 2HDM's, which have a $\tan\beta$ independent bound on the charged Higgs mass of $m_{H^+}\gtap 300\ \gev$, the bound on charged Higgs in Type I models is far weaker.  The constraint from the flavor changing process $B\rightarrow X_s\gamma$ allows a 100 GeV charged Higgs provided $\tan\beta\gtap 2.6$, which drops to $\tan\beta\gtap 1$ for $m_{H^+}\gtap 800\ \gev$ \cite{Mahmoudi:2009zx}. The presence of this additional light ($\sim 100 \, \gev$) charged scalar, whose masses can depend sensitively on the light Higgs mass, can lead to an additional contribution \cite{Posch:2010hx} to the decay width to $\gamma \gamma$, although we must be careful to apply the recent $t\rightarrow b H^+$, which plays an important constraint in the Type I case.  This constraint, as well, depends simply on simply the presence of a light charged Higgs with a significant coupling to the top (i.e., small $\tan \beta$).

In the simple model described above the mass spectrum, mixing angles, and couplings are all dependent on a few parameters in Higgs potential (\ref{eq:higgspot}).  Furthermore, there are constraints on the magnitude and sign of many of these couplings coming from the requirement of vacuum stability, perturbativity, and precision electroweak observables \cite{Posch:2010hx}.  The precision electroweak and heavy flavor constraints are weakened if the model is a sister Higgs model.  Finally, with a light charged Higgs the model is also subject to constraints from the bound on $t\rightarrow b H^+ (H^+\rightarrow \tau\nu)$ \cite{Aad:2012tj,CMS:2012cw}.  The LHC searches assume that the charged Higgs decays 100\% of the time to $\tau\nu$, whereas in the model in question this branching ratio is a function of charged Higgs mass and $\tan\beta$.  We take into account the competing modes of $c\bar{s}$ and $Wb\bar{b}$.  For values of $\tan\beta$ that saturate the $b\rightarrow s\gamma$ constraint the $\tau\nu$ BR is $\sim 70\%$ at low charged Higgs mass and  $\sim 10\%$ at 160 GeV.

\begin{figure}[t] 
   \centering
   \includegraphics[width=0.45\textwidth]{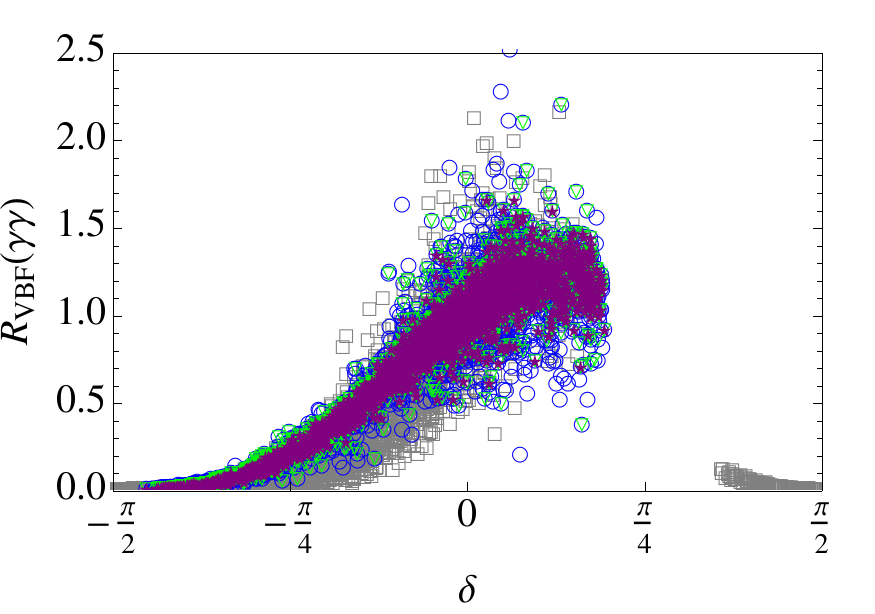} 
   \includegraphics[width=0.45\textwidth]{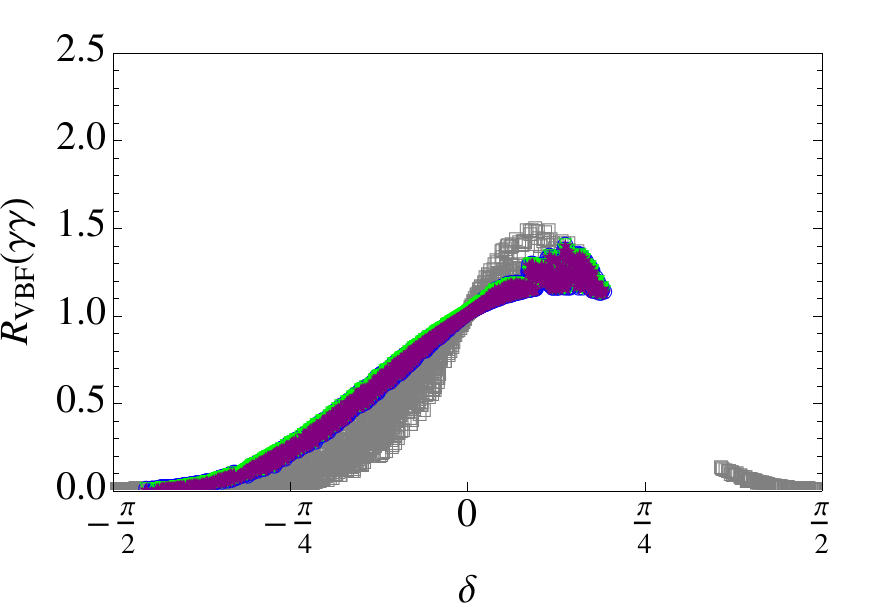} 
   \includegraphics[width=0.45\textwidth]{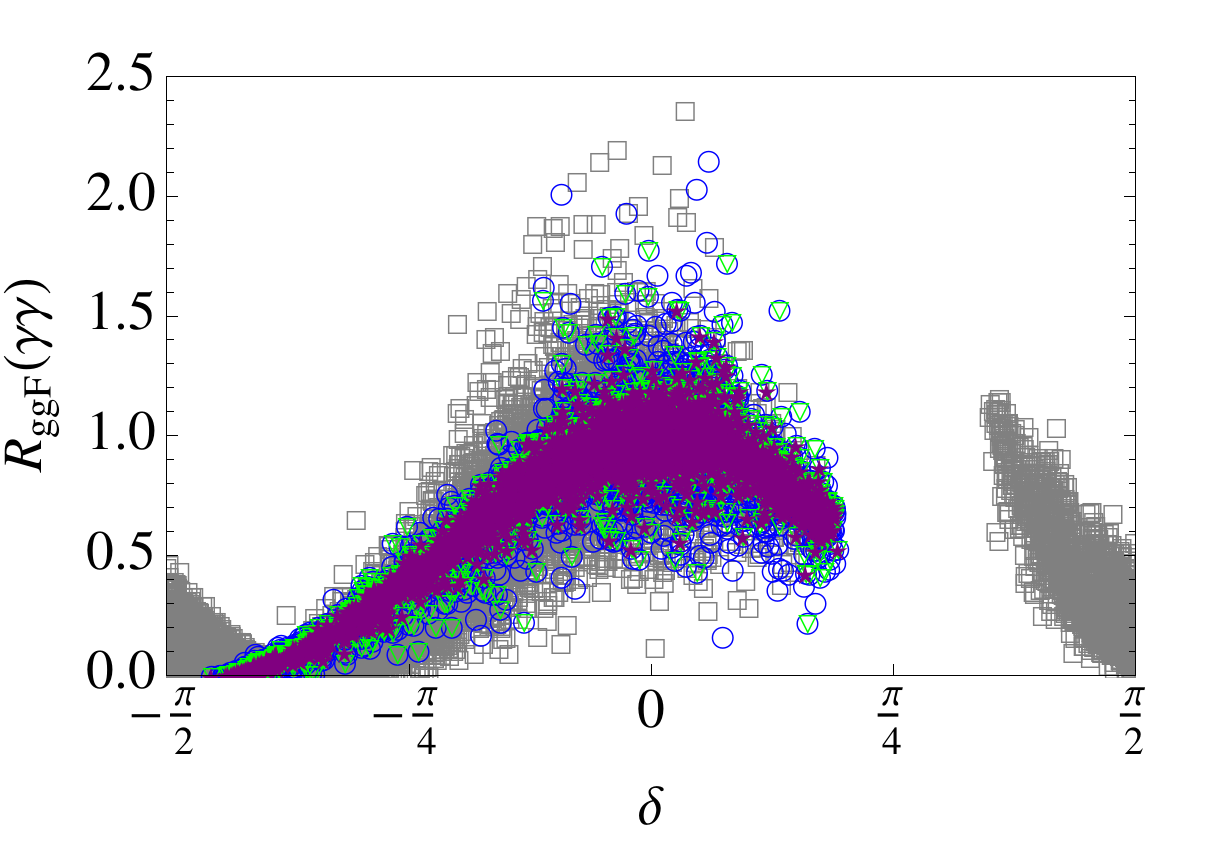} 
   \includegraphics[width=0.45\textwidth]{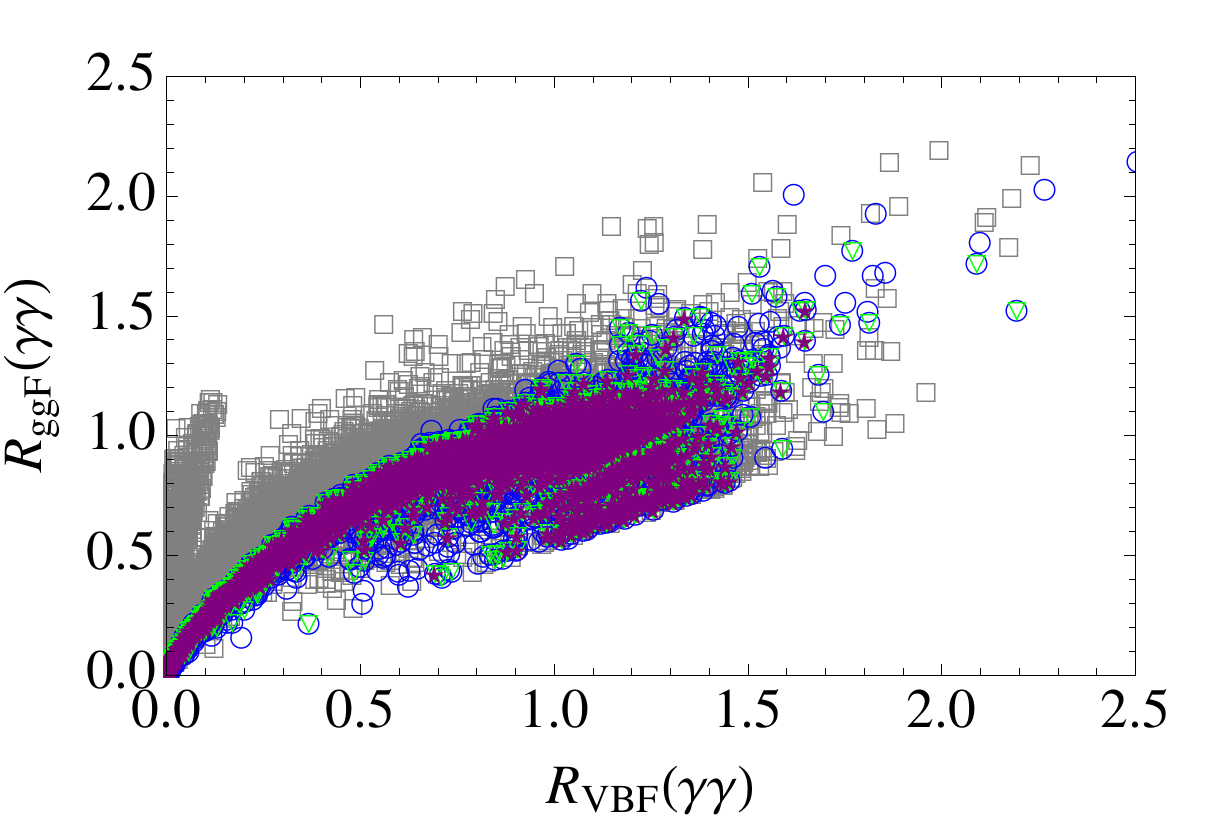} 
 \caption{ Results of a scan as described in the text, with the gray points ($\square$) corresponding to values of $\tan\beta$ failing the the $b\rightarrow s\gamma$ constraint, the blue points ($\circ$) satisfying this bound (but $\tan\beta<3$), but failing constraints on $S$ and $T$, the green ($\triangledown$) satisfying all these bounds, but failing the additional constraint from $t\rightarrow b H^+$, and the purple ($\star$) passing all these constraints. The upper right hand plot ignores the effects on $\Gamma_{\gamma\gamma}$ of the charged Higgs running in the loop, while all other plots include it.}
   \label{fig:scans}
\end{figure}

In order to determine the range of possibilities for the Type I/Sister Higgs model we carry out a scan over $\lambda_{2\ldots 7}$ and $m_{H+}$ fixing the other parameters by requiring the correct $W$ mass and the lightest CP even higgs is at 125 GeV.  We require that all couplings have magnitude less than 2, and consider charged Higgs masses in the range $[100,250]\ \gev$ and that the vacuum is stable \cite{Ferreira:2004yd,Ferreira:2009jb} i.e.
\begin{align}
\lambda_{1,2} & >0       & \lambda_3+\lambda_4-|\lambda_5| &>-\sqrt{\lambda_1\lambda_2} \\
\lambda_3 & >-\sqrt{\lambda_1\lambda_2}       & 2|\lambda_6+\lambda_7|  &<\frac{\lambda_1+\lambda_2}{2}+\lambda_3+\lambda_4+\lambda_5 ~.
\label{eq:vacstability}
\end{align}
We consider constraints from precision electroweak observables \cite{He:2001tp,Baak:2011ze}, limits from $t\rightarrow b H^+$ and $b\rightarrow s \gamma$. 
From this scan it is possible to see what regions of parameter space discussed above are obtainable in the Type I case. We are interested in both the physical signals ($R_i(X)$) as well as the equivalent, but more model-centric quantities $\tan \beta$, $\dcg$ and $\delta$. In Figure~\ref{fig:scans} we show a variety of distributions, specifically $R_{VBF}(\gamma\gamma)$ and $R_{gg}(\gamma\gamma)$ vs $\delta$, as well as $R_{VBF}(\gamma\gamma)$ vs $R_{gg}(\gamma\gamma)$. 

From these plots we can infer a number of results\footnote{The limited range of positive $\delta$ in Figure~\ref{fig:scans} is due to the fact that we only consider the range $0.5< \tan\beta < 5$.}. First, we see that sizeable effects can be had on both the VBF as well as inclusive signals. Looking over the overall scans, both signals can be much larger than the SM values, occasionally reaching rates $\sim 2\times$ SM.  At the same time, it is clear that experimental constraints, in particular $b\rightarrow s \gamma$ and $t\rightarrow b H^+$ constrain the parameter space considerably. Once these constraints are imposed, the range shrinks, with the upper range for the inclusive $\gamma\gamma$ signal topping at around 50\% and the VBF at around 60\% above the SM values. As we shall comment later, this makes extended sister Higgs models, where new charged states exist without these constraints, particularly interesting.
 
We should note that the points of maximal enhancement for the VBF and ggF signals occur for positive $\delta \sim \pi/8$. In this range, as we see in Figure \ref{fig:sigBRrel}, the $\bar \tau \tau$ signal can be suppressed below 40\% of the SM, while the $ZZ^*/WW^*$ signals remain similar to their SM values.

\begin{figure}[t] 
   \centering
   \includegraphics[width=0.5\textwidth]{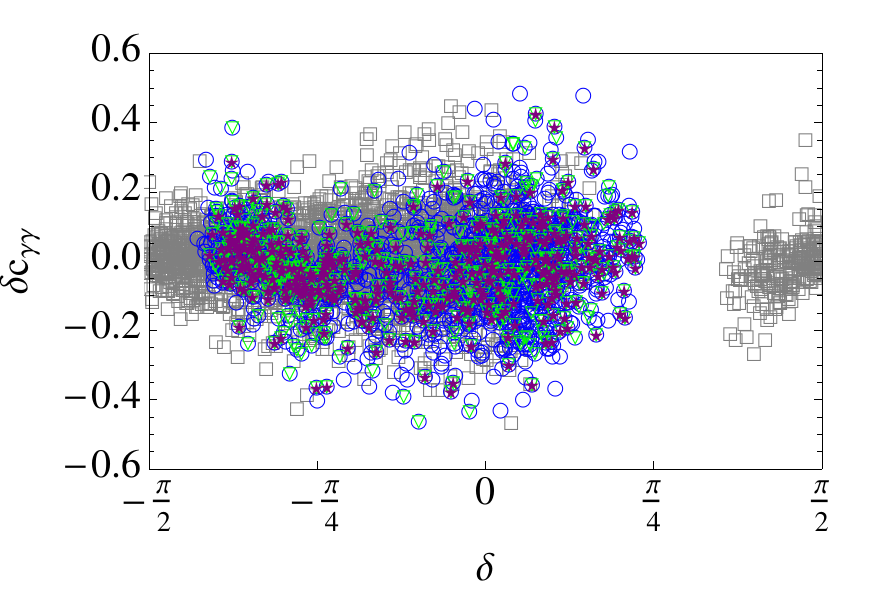} 
 \caption{The correction, $\delta c_{\gamma\gamma}$, to the diphoton width, $\Gamma_{\gamma\gamma}$, of the Higgs from the charged Higgs of the Type I 2HDM.  The color scheme is as in Figure~\ref{fig:scans}.}
   \label{fig:deltac}
\end{figure}

Another convenient way to look at these results is in terms of the contribution to $\dcg$, which we show in Figure \ref{fig:deltac}. Here we see that points satisfying all constraints can naturally yield $|\dcg| \approx 0.14$, which we then use as the envelope for the shaded region in Figure \ref{fig:sigBRrel}.

\begin{figure}[t] 
   \centering
   \includegraphics[width=0.45\textwidth]{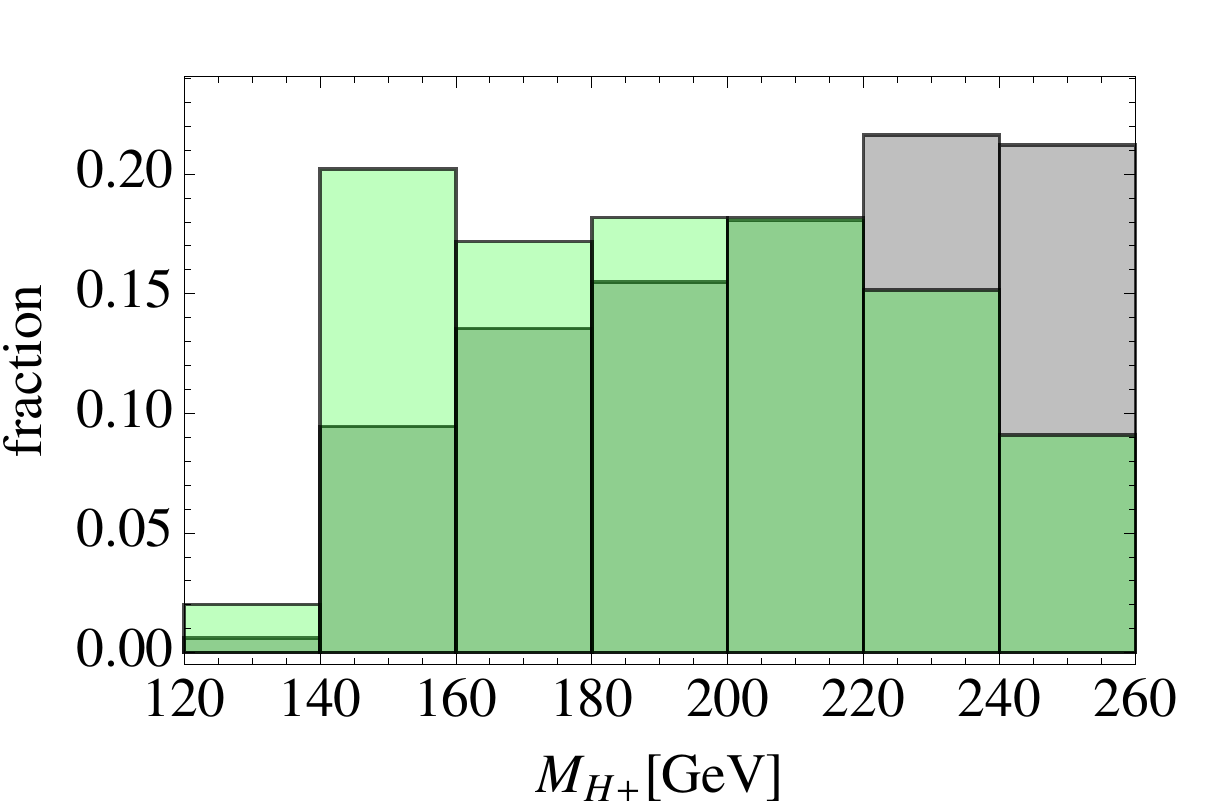} 
   \includegraphics[width=0.45\textwidth]{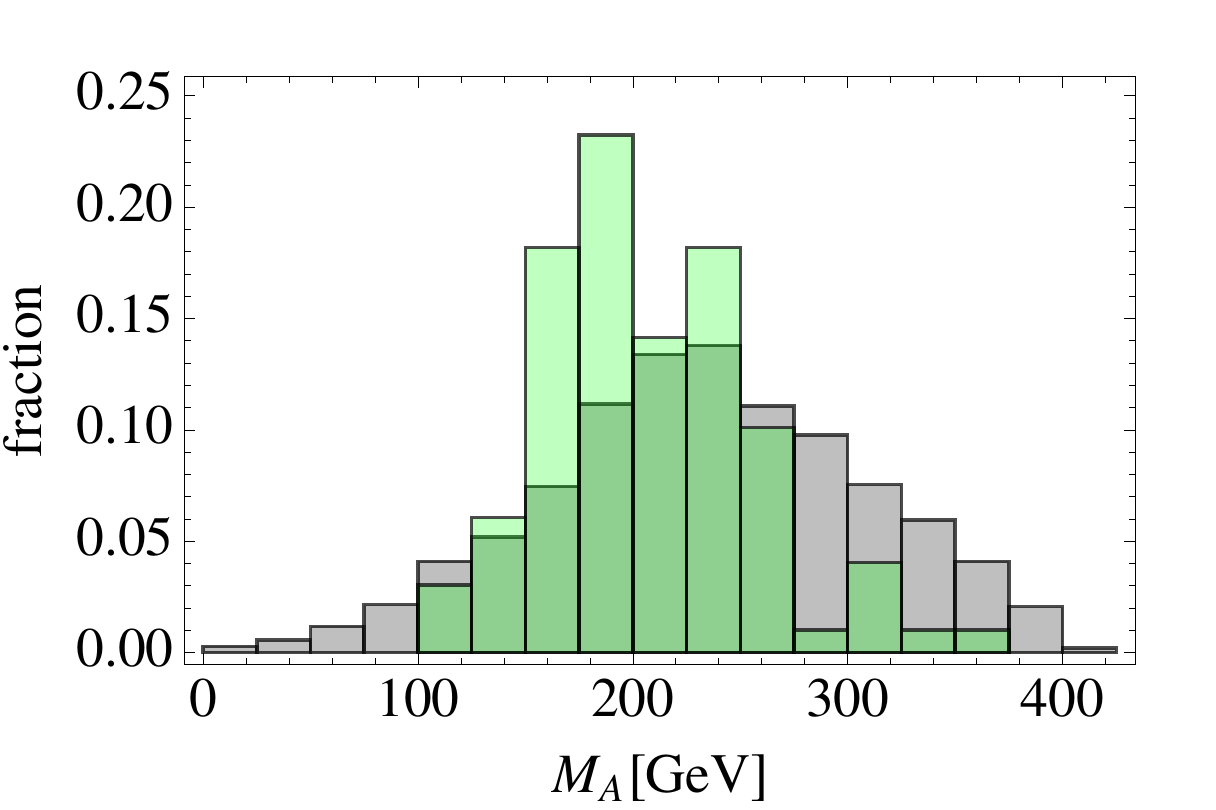}
   \includegraphics[width=0.45\textwidth]{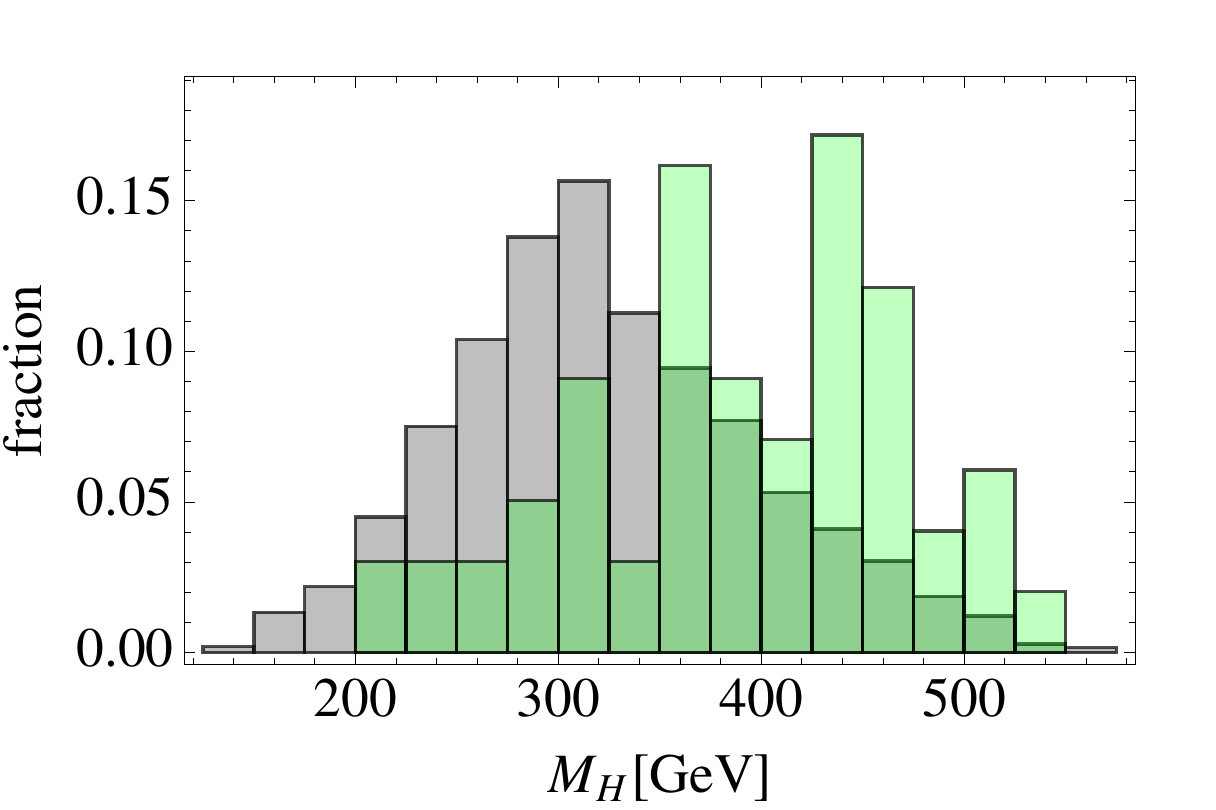}
 \caption{Distribution of heavy higgs masses in scan.  The gray applies all constraints, including perturbativity ($|\lambda|<2$) vacuum stability (\ref{eq:vacstability}), the constraints from $S$, $T$, $t \rightarrow b H^+$ and $b \rightarrow s \gamma$. The green has an additional requirement of $R_{VBF}(\gamma \gamma)>1.3$.}
   \label{fig:massdistributions}
\end{figure}

Also interesting are the masses for the other states in the theory.
The heavier Higgses in the theory have masses given by 
\bea
m_{A^0}^2 &=& \frac{2 m_{12}^2 - \frac{v^2}{2} (2\lambda_5\sin2\beta+\lambda_6(1+\cos2\beta)+\lambda_7(1-\cos2\beta))}{\sin2\beta}\\
m_{H^\pm}^2 &=& m_{A^0}^2+(\lambda_5-\lambda_4)v^2/2~.
\eea
The expression for the heavy CP even higgs mass is not illuminating.  As we carry out the scan we can also determine the masses of the heavier higgses, their distribution is shown in Figure~\ref{fig:massdistributions}.  As can be seen from Figure~\ref{fig:massdistributions} a light charged Higgs is a generic prediction of these models.  The coupling of the charged Higgses relative to those of the SM Higgs are
\bea
 c_f(H) &=& \frac{v g_{H f \bar f}}{m_f}= \sin\alpha/\sin\beta\approx -\delta - \cot\beta\\
 c_V(H) &=& \frac{v g_{H VV}}{2 m_V^2}=\cos(\beta-\alpha)\approx -\delta\\
 c_f(A) &=& \frac{v g_{A f \bar f}}{i m_f} = \cos\alpha/\sin\beta\approx 1-\delta \cot\beta~.
 \eea
The suppressed couplings of $H$, and the decreased branching ratio of $A$ to photons means that despite being light these states could, so far, have escaped detection at the LHC.  The strongest constraint here being in the limitation on $t\rightarrow b H^+$.

We show the results of the distributions of masses in Figure \ref{fig:massdistributions}. While we should not put any real stock in the actual distributions (there is no measure here), we can see that the imposition of a sizeable effect on the VBF signal pushes the charged Higgs to lighter values, along with $M_A$, but pushes the heavier Higgs mass $M_H$ to somewhat higher values.

\section{Realizations in Supersymmetry}
Within SUSY, a Type II 2HDM is generally considered to explain the origin of fermion masses (although see \cite{Dobrescu:2010mk}). The most natural realization of the physics described here is in the context of a ``sister'' Higgs model \cite{sh}, where the theory is extended by an additional $\Sigma_u, \Sigma_d$ into a four-Higgs doublet model. 

In the simplest scenario, the usual fermion masses are generated by Yukawa couplings to $H_u$ and $H_d$. One can yield a Type I 2HDM effective theory by integrating out $H_d$ with a $B \mu$ term (see the discussion in \cite{sh}), as well as one of the $\Sigma$ fields. For large $\vev{h_u}/\vev{h_d}$, one can describe much of the physics of the Higgses with a Type I model as we have here for $\vev{\Sigma}/\vev{h_u} \sim O(1)$. 

As argued in \cite{sh}, it can become quite natural to add new gauge charges to these sister Higgs fields under a sister group $G_s$, which can help yield a 125 GeV Higgs mass via an NMSSM-like quartic $\Phi H_u \Sigma_d$, where $\Phi$ is also charged under $G_s$. If such symmetries are non-Abelian, then there are additional states that will have quartics with the effective SM-like Higgs and contribute to its $\Gamma_{\gamma \gamma}$. In particular, the D-term quartics arising from $G_s$ can be reasonably large. These states are generally light due to the $D_s$ splitting in masses, which also gives the quartic the appropriate sign to enhance $\Gamma_{\gamma\gamma}$. These new states are not in general bound by heavy flavor constraints because they mix little, if at all, with the physical Higgs, and so their dominant couplings are just to the Higgs alone.

This is important because the dominant constraints on this model are from $b \rightarrow s \gamma$ and $t \rightarrow b H^+$, whice arise from $h-H$ mixing. In the presence of an additional $SU(2)$ partner field with a sizeable $h h^* H H^*$ coupling, but small or no mixing, large radiative effects may be present without such constraints. That is, in the context of these extended supersymmetric models, one might expect a set of sister Higgs fields that mix significantly with the SM-like Higgs, and an additional set that do not. Thus, we would expect both the tree-level effects as well as loop-level, but with the loop level generally being less constrained in these scenarios.

Finaly, considerations of Grand Unification motivate these theories to have $G_s$-charged colored fields (``G-quarks''). The scalar components of such fields would naturally be split and have an effective quartic $g_s^2/4 \cos^2(\beta -\delta) h h^* D_g^c D_g^{c*}$. For light $(\sim 150 \, \gev)$ G-squarks, large $(\alpha_s \sim 1)$ and $\tan \beta \sim 1$ an additional contribution comparable to (but generally smaller than) that of the charged Higgs can further be present.

We refer the reader to \cite{sh} for a broader discussion of these models and related issues.

\section{Conclusions}
In light of the recent discovery of a Higgs-like particle, there have been increased attention on models where the couplings of the Higgs are modified. We have argued that the inclusion of an additional Higgs doublet without couplings to fermions, but which participates in EWSB, can naturally provide a scenario where $c_V \approx 1$ and $c_f < 1$, in particular near $\tan \beta \sim 1$. At tree-level, this naturally leads to enhanced $\gamma \gamma$ signals in the VBF channel, suppressed $f \bar f$ decays and $WW^*/ZZ^*$ signals near the SM value or only moderately suppressed.

These tree-level effects (i.e., $\delta \ne 0$) are generally significant in the non-decoupling limit, with other scalar states at relatively low masses. The presence of these states leads to both important constraints from heavy flavor decays and electroweak precision. At the same time, these same states can induce significant radiative contributions to the $\gamma \gamma$ amplitude. This contribution then can lead to a further enhancement of the $\gamma \gamma$ signals (both inclusive and VBF), on top of a generally SM-like level of $WW^*/ZZ^*$, and a suppression of $f \bar f$ decays. This may provide a natural setting to explain the recent Higgs results, should the deviations from the SM become more statistically significant.

Such a new Higgs doublet is present minimally in a Type I 2HDM, or with the inclusion of a sister Higgs multiplet $\Sigma_u, \Sigma_d$ in supersymmetry. In such sister Higgs scenarios, the $\Sigma$ can be part of a multiplet, whose other components can serve to provide additional charged states to contribute radiatively to $\dcg$ without the stringent constraints from $b\rightarrow s \gamma$ and $t \rightarrow b H^+$. 

While detecting these new Higgs states may be challenging (as they lack the usual large $\tan \beta$ enhancements of a Type II model), it motivates a constrained fit of the different Higgs signals in terms of three parameters: $c_V, c_f$ and $\dcg$. If this overconstrained fit can be realized with such a simple parameterization, it would lend support to such scenarios.

Beyond this, in the context of a supersymmetric theory in particular, these sister Higgs fields would be expected to be accompanied by GUT partners, which are also charged under $G_s$. Such fields would have quartics that can further contribute to $\dcg$. These G-quarks and G-squarks would modify SUSY searches and may provide new signals to confirm a model of this type.

As more data arrive, it will become clearer whether the Higgs properties motivate new physics, but the scenario described here provides an economical explanation, while being naturally safe from electroweak precision, flavor and other constraints on new physics.

\section*{Acknowledgements}
We thank Graham Kribs, Joe Lykken and Dave Tucker-Smith for discussions. NW is supported by NSF grant \#0947827. Fermilab is operated by Fermi Research Alliance, LLC, under contract DE-AC02-07CH11359 with the United States Department of Energy.
\vskip 0.2in
{\bf \noindent Note added:} As this paper was being prepared for submission, \cite{Craig:2012vn} appeared, which similarly considers the tree-level effects of models with extended Higgs sectors in a general $\tan\beta-\sin\alpha$ parameter space.

\bibliography{typeI}
\bibliographystyle{apsrev}

\end{document}